\documentstyle[seceq,epsf,wrapft]{ptptex}

\title{Role of
quark-quark correlation in baryon structure and non-leptonic weak
transitions of hyperons }

\author{
Emiko \textsc{ Hiyama}$^{1,}$\footnote{
 E-mail: hiyama@post.kek.jp},
 Katsuhiko \textsc{ Suzuki}$^{2,}$\footnote{E-mail:
 ksuzuki@la.numazu-ct.ac.jp},
 Hiroshi \textsc{ Toki}$^{3,}$\footnote{E-mail:
 toki@rcnp.osaka-u.ac.jp} and\\
Masayasu \textsc{ Kamimura}$^{4,}$\footnote{
 E-mail: kami2scp@mbox.nc.kyushu-u.ac.jp}
}

\inst{
$^1$Institute of Particle and Nuclear Studies, High Energy
Accelerator Research Organization (KEK), Oho, Tsukuba, Ibaraki
305-0801, Japan\\
$^2$Numazu College of Technology, Numazu 410-8501, Japan\\
$^3$Research Center for Nuclear Physics (RCNP), Osaka University,
Ibaraki, \\Osaka 567-0047, Japan\\
$^4$Department of Physics, Kyushyu University, Fukuoka 812-8581,
Japan }

\recdate{
}

\abst{
We study the role of quark-quark correlation in the baryon
structure and, in particular,  the hyperon non-leptonic weak
decay, which is sensitive to the correlation between quarks in the
spin-0 channel. We rigorously solve non-relativistic three-body
problem for SU(3) ground state baryons to take into account the
quark-pair correlation explicitly. With the suitable attraction in
the spin-0 channel, resulting static baryon properties as well as
the parity conserving weak decay amplitudes agree with the
experimental values. Special emphasis is placed also on the effect
of the SU(6) spin-flavor symmetry breaking on the baryon
structure. Although the SU(6) breaking effects on the local
behavior of the quark wave functions are considerable due to the
spin-0 attraction, the calculated magnetic moments are almost the
same as the naive SU(6) expectations. }

\begin{document}

\maketitle

\section{Introduction}

Properties of light baryons have been extensively studied by
various models based on the
constituent quark picture, in which
the constituent quarks are assumed to be
identified with quasi-particles of non-perturbative QCD vacuum.
Their results for static hadron properties are consistent
with experiments including applications for the two nucleon
systems,
although this model involves several adjustable parameters.
Despite the success of this approach, it is not clear whether or
not the constituent quark model correctly describes the quark
distributions in the baryons, and further provides the
non-leptonic weak hyperon decay with $\Delta I = 1/2$ rule.

The $\Delta I = 1/2$ rule implies dominance of the $\Delta I =1/2$
transitions and strong suppression of $\Delta I = 3/2$ process in
the non-leptonic hyperon decay\cite{Review}. This empirical rule
does not originate from the theory of the weak interaction itself,
and therefore one needs an explanation of some dynamical origin.
As we will show in section 2, the non-leptonic weak decay of
hyperons is reasonably described by the soft-pion theorem. It
leads to the baryon pole approximation which can reproduce
relative magnitudes of various hyperon decays and $\Delta I=1/2$
rule. Within this approximation, the weak decay takes place as the
two-quark transition process where a $us$-pair in the initial
hyperon with their total spin 0 changes to a spin-0 $ud$-pair in
the final state baryon. However, if one calculates its matrix
element using the constituent quark model, the absolute value of
the amplitude is about a half of the experimental data at
most\cite{model}\footnote{It may be possible to reproduce the
experimental values by artificially taking a smaller baryon
radius.  Such a procedure cannot be justified, because the baryon
radii should be also chosen to be consistent with experiments.}.
We emphasize here, because of the heavy $W$-boson mass, the weak
matrix elements are quite sensitive to the short range quark-quark
correlations. Consequently, the failure of the constituent quark
model to describe the non-leptonic weak transition may suggest the
lack of the quark correlation in the $s=0$ channel. Several
works\cite{Bando,Diquark} also suggest the importance of the
short-range quark correlation on the $\Delta I = 1/2$ decay. In
particular, it is known that the parity-conserving $\Sigma ^+ \to
n \pi^+$ decay is free from factorization and penguin
contributions, and simply dominated by the two-quark transition
process in the baryon pole term . Therefore, this decay mode gives
a crucial constraint on the strength of the quark-quark
correlation.

On the other hand, it was recently pointed out from both
theoretical\cite{Instanton,QSR} and phenomenological\cite{diq-exp}
points of view, there exists a strong correlation between quarks
in the $s=0$ channel, corresponding to the correlation between the
quark-antiquark spin-0 pair which forms the pion as the highly
collective state. The simple constituent quark model has never
incorporated such a quark-quark correlation properly.

In this paper we try to clarify the role of the quark-quark
correlation in the baryon structure and hyperon non-leptonic weak
decays by using the constituent quark model. We assume the short
range spin-dependent correlations between quarks together with the
confinement force, and calculate the baryon masses and other
static properties. In order to deal with the spin-dependent
correlation in the correct and systematic way, we must  rigorously
solve the three-body problem. For this purpose, we adopt the
Gaussian expansion method (GEM) for few-body systems,
 which has been developed by the
two of the present authors (E.H. and M.K.) and their collaborators
\cite{Hiyama03,Kami88,Kame89,Hiyama95,Hiya96,Hiya00} 
(see Ref. 9 for a review). We assume
the isospin symmetry between $u$ and $d$ quarks, and solve the
three-body problem without further approximations or assumptions.
Our calculations do not rely on the SU(3) flavor symmetry, and
hence the strange quark is distinguished from light $u,d$ quarks.
To our knowledge, this work is a first attempt to study the
non-leptonic hyperon decay and other hadron properties
consistently in the framework of the constituent quark model by
taking into account the quark-quark correlation.

So far the constituent quark model with the confining force and
the perturbative gluon exchange provides reasonable results for
masses and radii, but fails in reproducing the hyperon weak
decay\cite{model}. After introducing the suitable quark-pair
correlation, one does not know {\it a priori} whether or not the
constituent quark model can explain both standard properties and
at the same time non-leptonic weak decay matrix elements. One of
the main purposes of this paper is to make this issue clear by
solving the non-relativistic three-quark problem explicitly.

We also focus on the SU(6) breaking effects on baryon properties.
Introduction of the spin-dependent correlation naturally spoils
the SU(6) spin-flavor symmetry which is known to work well
for {\it e.g.}~the baryon magnetic moments.
Hence, we shall calculate the magnetic moments to estimate
the SU(6) breaking effects.
We discuss how the SU(6) breaking affects the baryon properties,
and point out that this symmetry is still useful for the static baryon
properties, although local behavior of the quark wave function
considerably departs from the SU(6) symmetric limit.

This paper is organized as follows. In section 2, we introduce the
effective weak interaction which includes the renormalization
group improved QCD corrections.  We show the relevant formulae
within the pole approximation, and define a set of matrix
elements. In section 3 we construct the non-relativistic potential
model which incorporates the confinement force and the quark
correlation. Our numerical procedure to solve the three-body
problem is described in detail here. In section 4 we calculate the
baryon masses, radii, and the non-leptonic weak transition
amplitudes. Comparison with experiments will be made there. We
also evaluate the magnetic moments of  the SU(3) baryons in order
to clarify the SU(6) breaking effects in section 5. Final section
is devoted to the summary and discussions.

\section{Calculation of weak matrix elements}

We write the low energy effective weak interaction Hamiltonian
density\cite{Hamiltonian};
\begin{equation}
{\cal H}_{W} (x) ={{G_F\sin \theta \cos \theta } \over {\sqrt 2}}
\sum\limits_i^{} {c_i(\mu ^2)\;}O_i (x) + \mbox{h.c.} \quad ,
\label{whamiltonian}
\end{equation}
where
\begin{eqnarray*}
O_1&=&[\bar u\gamma _\mu (1-\gamma _5)s][\bar d\gamma ^\mu (1-\gamma _5)u]
\\
O_2&=&[\bar d\gamma _\mu (1-\gamma _5)s][\bar u\gamma ^\mu (1-\gamma _5)u]
\; \\
O_3&=&[\bar d\gamma _\mu (1-\gamma _5)s]\sum\limits_{q=u,s,d}
{[\bar q\gamma ^\mu (1-\gamma _5)q]}\\
O_4&=& \sum\limits_{q=u,s,d} {[\bar q\gamma _\mu (1-\gamma _5)s]}
[\bar d\gamma ^\mu (1 - \gamma _5)q] \\
O_5&=&[\bar d\gamma _\mu (1-\gamma _5)s]\sum\limits_{q=u,s,d}
{[\bar q\gamma ^\mu (1+\gamma _5)q]}  \\
O_6&=& \sum\limits_{q=u,s,d} {[\bar q\gamma _\mu (1-\gamma _5)s]}
[\bar d\gamma ^\mu (1+\gamma _5)q] .
\end{eqnarray*}
This effective weak Hamiltonian can be obtained by integrating out over
$W$-boson
degrees of freedom in the Standard Model.
Current-current
operators $O_1, O_2$ give dominant contributions in our case, and
$O_3 \sim O_6$ provide so called penguin contributions.
The coefficients $c_i(\mu^2)$ of the Hamiltonian
get QCD radiative corrections calculated by the
renormalization group technique, and thus depend on the
scale.  We take the  scale $\mu^2 = 1 {\rm GeV}^2$
 as a typical scale of light hadrons, and use values of $c_i$
given in ref. 15).

Our task here is to evaluate the matrix element
$\langle B_f \; \pi^a | {\cal H}_{W} (0) |  B_i \rangle $
for the strangeness changing
process
$B_i \to B_f + \pi^a$.
We note that the factorization as well as the penguin
contributions are too small to reproduce the $\Delta I=1/2$ amplitudes.
QCD radiative corrections to the effective weak Hamiltonian
tend to increase the $\Delta I=1/2$ amplitudes, but effects are
not enough.

Analysis based on the chiral dynamics of low energy QCD is
suitable to deal with the strongly interacting pion-nucleon
system. With the help of the soft pion theorem, one can rewrite
the transition matrix element as
\begin{eqnarray}
\left\langle {B_f(p_f)\;\pi ^a(q)|{\cal H}_{W}(0) \,|B_i(p_i)} \right\rangle
& = & \int {d^4x}\,e^{-iqx}(-q^2+m_\pi ^2)\left\langle
{B_f\;|T\left\{ \pi ^a(x),{\cal H}_{W} (0) \right\}\,|B_i} \right\rangle
 \nonumber \\
& =&{{-1} \over {f_\pi }}\left\langle {B_f\;|
\left[ i\int {d^3x\,} A_a^0(x),{\cal H}_{W}(0) \right] \,|B_i} \right\rangle
\nonumber  \\
&&\hspace{-0.5 cm} +{{iq_\mu }
\over {f_\pi }}\int {d^4x}\,e^{-iqx}\left\langle
{B_f\;|T\left \{  A_a^\mu (x),{\cal H}_{W}(0) \right \} \,|B_i}
\right\rangle.
\label{softpion}
\end{eqnarray}
The first term is called the commutator term which gives the
parity violating S-wave amplitudes.  The second term expresses the
baryon pole contribution by inserting the intermediate baryon states
between $A_\mu^a (x)$ and ${\cal H}_W (0)$, and contributes to both
parity conserving and violating amplitudes, as illustrated in Fig. 1.
\begin{figure}
\epsfxsize = 14cm   
\centerline{\epsfbox{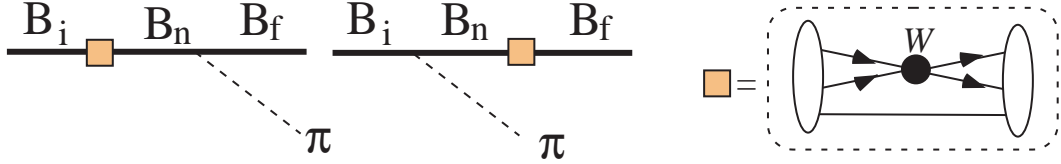}} \caption{Baryon pole contributions
to the non-leptonic hyperon decay. Thick and thin solid lines
denote the baryon and quark, respectively.  The pion is depicted
by the dashed line. The weak transition between baryon states is
shown by the filled box, which expresses the two quark weak
transition process $us \to ud$ shown in the right hand side.}
\label{weakf}
\end{figure}
Here, the initial hyperon $B_i$ changes to an intermediate state
baryon $B_n$ by the weak interaction and then $B_n$ emits the pion
to produce the final state $B_f$(and vice versa). In this work, we
concentrate on the parity conserving P-wave amplitudes to
investigate the quark correlation inside the ground state
baryons\footnote{The parity violating amplitudes are not adequate
for the study of the quark correlation selectively, since the
contributions from the commutator, penguin, factorization, and the
baryon pole are all comparable.}. For example, the parity
conserving amplitude for $\Lambda^0 \to n + \pi ^0$ is given by
\begin{eqnarray}
B (\Lambda ^0 \to n + \pi^0) &&  \nonumber \\
&&\hspace{-3.5 cm} ={{M_N+M_\Lambda } \over {f_\pi }}\left[ {G_{nn}^{\pi 0}
{1 \over {M_\Lambda -M_N}}\left\langle {n\left|{\cal H}_{PC}
  \right|\Lambda }
\right\rangle +\left\langle {n\left| {\cal H}_{PC} \right|\Sigma
^0} \right\rangle {1 \over {M_n-M_\Sigma }}G_{\Lambda \Sigma
}^{\pi 0}} \right].
 \label{lambdapole}
\end{eqnarray}
where $\left\langle {n\left|{\cal H}_{PC}
  \right|\Lambda } \right\rangle $ and
$\left\langle {n\left| {\cal H}_{PC} \right|\Sigma ^0}
\right\rangle $ are the matrix elements  between
appropriate baryon states with ${\cal H}_{PC}$ being the parity
conserving part of the weak Hamiltonian defined in the Appendix.
$G_{B \; B' }^{\pi a}$ denotes the axial vector coupling constant
which gives a probability for the pion emission $B \to B' +
\pi^a$. Here, we consider only the ground state baryon octet as
the intermediate states. Formulae for other hyperon decays are
found in the Appendix.


The axial vector coupling constants are rather well-known quantities
from experiments.
We adopt the values of $G_{B \; B'} ^{\pi a}$ obtained by the SU(3)
parameterization for the existing experimental data, since the
axial vector coupling seems to be insensitive to the quark correlation
due to its one-body operator structure.
Therefore, we are now in the position to determine the matrix elements
of the
weak Hamiltonian, $\langle n | {\cal H}_{PC} | \Lambda \rangle$ and
$\langle p | {\cal H}_{PC} | \Sigma^+ \rangle$\footnote{In this paper,
we restrict
ourselves to study only $\Lambda, \Sigma$ hyperon decay.
Results including other hyperons
will be published in subsequent publication.}.
We recall that the quark models such as Isgur-Karl Harmonic Oscillator
model\cite{Isgur} or
MIT bag model\cite{MIT} give much smaller values for these
matrix elements than the experimental data\cite{model,Bando}.
It is instructive to rewrite the weak Hamiltonian ${\cal H}_{PC}$ in the
non-relativistic limit in the coordinate space as
\begin{equation}
{\cal H}_{PC} = \frac{G_F \sin \theta \cos \theta }
{\sqrt{2}} (c_1 O_1^{NR} +  c_2 O_2^{NR} ) \;, 
\end{equation}
\begin{equation}
O_1^{NR} =  - O_2 ^{NR} =
a_d^\dagger \,  a_u^\dagger (1 - \vec \sigma_u \cdot \vec \sigma_s)
\, \,  \delta^{(3)} (\vec r_{us}) \, \, a_u  a_s \; ,
\label{opew}
\end{equation}
where $a_i$, $ a_i^\dagger$ are annihilation and creation
operators of quarks with flavor $i$.  Presence of the
spin-projection operator $(1 - \vec \sigma_u \cdot \vec \sigma_s)$
in $O_1^{NR}, O_2^{NR}$ tells us that only the $(us)$ pair with
their total spin being 0 can contribute to the weak decay process,
namely, the weak transition is generated by the two body process
between spin-0 quark pairs; $(us)^0 \to (ud)^0$. The isospin of
the initial $(us)^0$ pair is $1/2$, and the final $(ud)^0$ has the
isospin-0 due to the antisymmetrization. Thus, this process
guarantees the $\Delta I=1/2$ dominance in the non-leptonic
hyperon decays as pointed out long ago\cite{Review}. Now it is
clear that this decay amplitude is very sensitive to the
correlation of the spin-0 quark pair in the baryons\cite{Diquark}.
The standard constituent quark model never incorporates such a
correlation properly. However, in fact, some fundamental studies
on non-perturbative QCD\cite{Instanton,QSR} suggest that there
exists the strong attractive correlation for the quark-quark pair
with $s=0$. These considerations naturally lead us to study the
quark structure of baryons by taking into account the attractive
correlation which could enhance the weak decay amplitudes.

\section{Constituent quark model for baryons with spin-dependent
correlation}

Our purpose here is to construct the quark model to deal with the
quark pair correlations and thus account for the non-leptonic weak
decay. Although there are several efforts from the lattice QCD
simulation and phenomenological analysis, our knowledge of the
interaction between light quarks is still far from complete
understanding. As we have discussed in the introduction, the
instanton liquid model of the non-perturbative QCD vacuum provides
a spin-dependent correlation between the quark-pair.  Due to the
finite spatial size of the instantons, typically
0.35fm\cite{Instanton2}, such a spin-spin interaction should not
be point-like but has a finite range.  The constituent quarks are
also assumed to have their internal structure. Hence, we introduce
the Gaussian shape spin-dependent force which acts on only the
quark-quark pair with $s=0$. We neglect possible flavor dependence
of the potential. Other spin-dependent pieces like the spin-orbit
and tensor interactions are neglected for simplicity, because we
shall clarify the effects of the spin-spin interaction on the
non-leptonic weak decay amplitudes.

On the other hand,
we take the two-body Harmonic Oscillator potential as the confinement force,
since the analytical solutions of the three-body system are well
known\cite{Isgur}.
In order to check accuracy of our numerical calculations, we can
compare analytical results with ours, when we turn off the
spin-dependent interaction.
Choice of the HO potential is advantageous for us to develop
our numerical procedure in this paper, but it can be
easily improved in a more realistic way.

Finally, we phenomenologically introduce the effective
Hamiltonian which includes the confinement force $V_C$ and the
spin-dependent part $V_S$ as
\begin{eqnarray}
 H &=&\sum_i {{{{\bf p}_i^2} \over {2m_i}}}
- T_G + V_{C} + V_{S} + V_0 \;,
\end{eqnarray}
\begin{eqnarray}
V_{C} =\sum_{i<j} {{1 \over 2}K \left( {\bf x}_i- {\bf x}_j
\right)^2},
\end{eqnarray}
\begin{equation}
V_{S}=\sum_{i<j} {{C_{SS}} \over {m_i m_j}} \mbox{exp}\left[
-\left( {\bf x}_i-{\bf x}_j \right)^2
 / \beta ^2 \right] \quad (\mbox{spin=0 pair})
\end{equation}
and $V_S=0$ for spin=1 pair.
%
%
%
Here, $m_i$, ${\bf x}_i$ and ${\bf p}_i$ are the mass, coordinate
and momentum of the $i$-th constituent quark,  and $K, C_{SS},
\beta$ are the model parameters which are taken to be common for
all the baryons concerned. $T_G$ is the c.m. kinetic energy and
$V_0$ is the constant parameter which contributes to the over all
shift of the resulting spectrum and is chosen to adjust the energy
of the lowest state to the nucleon mass. The quark masses are
taken to be $m_u = m_d = 330$ MeV and $m_s = 500$ MeV.

Using this Hamiltonian, we shall solve non-relativistic
three-body problem rigorously.
We assume only the isospin symmetry between up and down quarks.
The quark wave functions are constructed by the antisymmetrization
without invoking any further approximations or assumptions.
We note that the SU(6) spin-flavor symmetry
is broken within our
formalism because of the spin-dependent correlation.
Namely, the spatial part of the total wave function
is dependent on the spins and isospins of the three quarks
and is expanded in terms of a number of basis functions
so as to describe the spin-dependent short-range correlations.

Since $N, \Delta$ and $\Omega^-$ are composed of three quarks
having the same isospin and the other baryons
 $\Lambda, \Sigma, \Sigma^*
\Xi$ and $\Xi^*$ are not, we introduce two different types of
total wave functions for the two cases.

\subsection{ Wave functions of  $N, \Delta$ and $\Omega^-$}

According to the Gaussian expansion method
\cite{Hiyama03,Kami88,Kame89,Hiyama95,Hiya96,Hiya00},
we consider three rearrangement Jacobian coordinates of Fig. 2
and refer them as channels $c=1$ to 3;
here, ${\bf r}_k={\bf x}_i - {\bf x}_j$ and
${\bf R}_k = {\bf x}_k -({\bf x}_i + {\bf x}_j)/2$
for  the cyclic permutations of $(i, j, k)$.
\begin{figure}
\epsfxsize = 12cm   
\centerline{\epsfbox{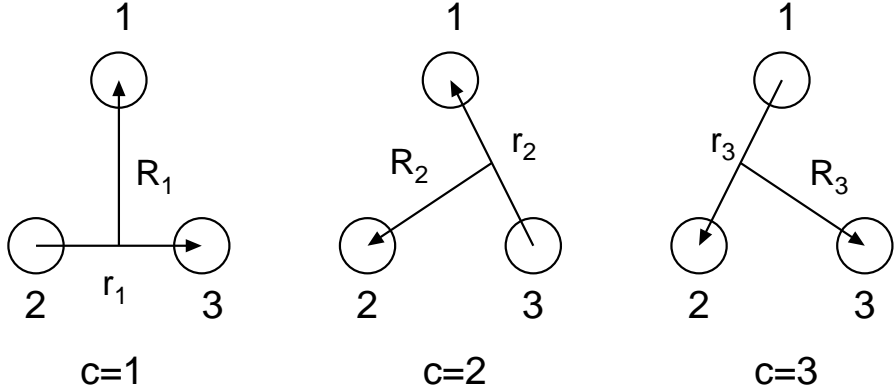}}
\caption{Rearrangement Jacobian coordinates for $N, \Delta,$ and $\Omega$}
\label{rjc1}
\end{figure}
We first construct three-body  basis functions
for the spin, isospin and spatial part of the channel $c=k$
 with $J, M$ (total spin and its $z$-component) and
$T, T_z$ (total isospin and its $z$-component) as
\begin{eqnarray}
\Phi _{JMTT_z,\:\xi}^{(c=k)} &=&
\bigg[
\Big[ [ \chi _\frac{1}{2}(i)\: \chi _\frac{1}{2}(j) ]_s
\chi _\frac{1}{2}(k) \Big]_S
\Big[ \phi _{n l}({\bf r}_k)\: \psi _{NL}({\bf  R}_k) \Big]_I
\bigg]_{J M} \nonumber \\
& & \times
\Big[ \left[ \eta _{\tau}(i)\: \eta _{\tau}(j) \right]_t
\eta _{\tau}(k) \Big]_{T T_Z} \; , \qquad 
\xi \equiv \{s,S,n,l,N,L,I,t \} \quad
\label{base}
\end{eqnarray}
with  the isospin of quarks, $\tau$, to be
\begin{eqnarray}
\tau \; = \left\{ \begin{array}{ll} 
 \;\;  \frac{1}{2} \;\;(u, d) \;\;\;\; \mbox{for $N, \Delta$} \\
  \!\! \;\;\;  0 \;\;(s)     \;\;\;  \mbox{for $\Omega^-$}\; . 
\quad \quad \quad \quad \quad \quad \quad 
\quad \quad \quad  \end{array} \right.
\end{eqnarray}
Here, $\eta _{\tau \tau_z}(i)$ is the isospin function
of the $i$-th quark,
$t$ denoting the isospin of each quark-pair.
$\chi _{\frac{1}{2}m}(i)$ is
the spin function of the $i$-th quark.
$s$ and $S$ denote the intrinsic spin of each quark-pair and
three quarks, respectively.
The numbers $n$ and $l$ ($N$ and $L$) specify respectively
the radial and  angular-momentum excitations with respect
to the Jacobian coordinates ${\bf r}_c$ (${\bf R}_c$),
and $l$ and $L$ are coupled to
the total orbital angular momentum $I$.
Explicit form of the spatial functions
$\phi _{n l}({\bf r}_k)$ and $\psi _{NL}({\bf  R}_k)$
will be discussed below. The three-body center-of-mass motion does not appear in our framework.

The totally antisymmetric basis function with the quantum number set
$\xi$, $\Phi_{JMTT_z, \;\xi}$,
 is obtained by the {\it equal-weight} superposition
of $\Phi_{JMTT_z, \;\xi}^{(c)}$ over
$c=1-3$, multiplied by the color singlet wave function and
posed by the Pauli restriction $1+ s + 2\tau + t + l =  even $ :
\begin{equation}
\Phi _{JMTT_z, \:\xi}= \; \Phi(\mbox{color singlet})
     \sum_{c=1}^{3} \Phi_{JMTT_z, \;\xi}^{(c)} \; .
\label{antibase}
\end{equation}
The totally antisymmetric property of
the basis functions $\Phi_{JMTT_z,\xi}$ is explicitly seen
by interchanging the particle numbers of any pair.

The total  wave function, $\Psi _{JMTT_z}$,
is given as a sum of these basis functions:
\begin{equation}
\Psi _{JMTT_z}=\sum_\xi  A_{\: \xi}\;  \Phi_{JMTT_z, \;\xi}\; .
\label{total}
\end{equation}
This form is the most general one of
the totally antisymmetric three-quark functions
of   $N, \Delta$ and $\Omega^-$ (and their spatially excited states).
The coefficients $A_{\:\xi}$ are to be determined
by solving  the Schr\"{o}dinger equation
\begin{equation}
        ( H - E )\: \Psi _{JMTT_z} = 0
\end{equation}
with the Rayleigh-Litz variational principle.

\subsection{ Wave functions of  $\Lambda, \Sigma, \Sigma^*,
\Xi$ and $\Xi^*$}

In this case, in order to construct totally antisymmetric
wavefunctions, we consider according to the Gaussian expansion
method \cite{Hiyama03,Kami88,Kame89,Hiyama95,Hiya96,Hiya00}, the
nine  rearrangement Jacobian coordinates of Fig. 3 $(\gamma=1-3, \;
c=1-3)$ in which a particle (illustrated by a double circle) is
$s$ quark and the other two are $u, d$ quarks for $\Lambda,
\Sigma, \Sigma^*$, and the situation is opposite for $\Xi$,
$\Xi^*$. The channel name $\gamma$ indicates the particle number
of the sole, different quark (double circle), whereas the channel
name $c$ denotes the particle number which the Jacobian coordinate
{\bf R} points. In order to make the coupling scheme of the spins
and isospins of the three quarks as visual as possible, we always
place the sole, different quark as the thirdly (lastly) coupled
particle in the spin-isospin space (but not always in this order
in the coupling scheme of the coordinate space).

The three-body basis function for the spin, isospin and
spatial part of the channel $\gamma, c$ is given by
\begin{eqnarray}
      \Phi_{JMTT_z,\: \xi}^{(\gamma, \: c)} & = &
       \bigg [ \Big[ \big[ \chi_\frac{1}{2}(\alpha)
       \chi_\frac{1}{2}(\beta) \big]_s \chi_\frac{1}{2}(\gamma)
\Big]_S
       \Big[
   \phi_{n l}({\bf r}_{\gamma, c}) \psi_{N L}({\bf R}_{\gamma.c})
\Big]_I
     \bigg]_{JM}  \nonumber \\
     & & \times \bigg[ \Big[ \eta_{\tau}(\alpha) \eta_{\tau}
(\beta) \Big]_t
       \eta_{\tau_\gamma}(\gamma) \bigg]_{T T_z} \;, \qquad  
\xi \equiv \{s,S,n,l,N,L,I,t \} \quad
\end{eqnarray}
where $\alpha, \beta, \gamma$ are given by the cyclic permutations.
The isospins $\tau$ and $\tau_\gamma$ are defined as
\begin{eqnarray}
\tau = \left\{ \begin{array}{ll} 
 \;\;  \frac{1}{2} \;\;(u, d) & \;\;\mbox{for $\Lambda, \Sigma, 
   \Sigma^*$} \\
   \!\!\;\;\;\;  0 \;\;(s)  &  \;\;   \mbox{for $\Xi, \Xi^*$}\; , 
\end{array}\right.
 \qquad
 \tau_\gamma = \left\{ \begin{array}{ll}  
   \!   0 \;\;(s) \;\;\; &  \mbox{for $\Lambda, \Sigma, \Sigma^*$} \\
    \;\;  \frac{1}{2} \;\;(u, d) &    \mbox{for $\Xi, \Xi^*$}\; , 
 \end{array} \right. 
\end{eqnarray}
\begin{figure}
\epsfxsize = 10cm   
\centerline{\epsfbox{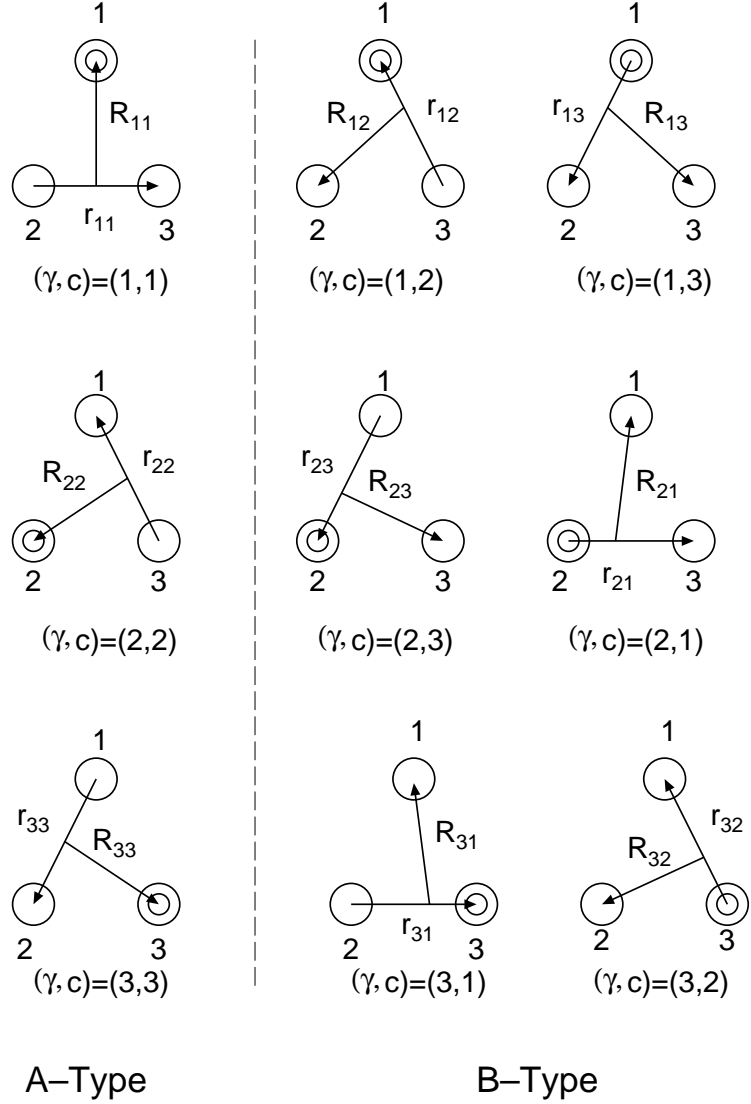}}
\caption{Rearrangement Jacobian coordinates for $\Lambda, \Sigma,
\Sigma^*, \Xi,$ and $\Xi^*$}
\label{rjc2}
\end{figure}
Using these basis functions, we can construct two types of
totally antisymmetric three-quark basis functions:

(i) A-type: {\it equal-weight} superposition of the three basis functions
with $\gamma=c=1-3 \:$ (cf. the left-most column of  Fig. 3),
multiplied by  $\Phi$(color singlet)
and posed by the Pauli restriction
$1+ s + 2\tau + t + l =  even $ for the quark pair
having the same isospin (strangeness):
\begin{equation}
     \Phi_{JMTT_z, \:\xi}^{\rm (A)}  = \: \Phi \mbox{(color singlet)}
     \sum_{\gamma=c=1}^{3}
  \Phi_{JMTT_z, \:\xi}^{(\gamma, c)}  \;
.
\end{equation}

(ii) B-type: {\it equal-weight} superposition of the six basis functions
with $\gamma\ne c=1-3 \:$ (cf. the middle and the right columns
of  Fig. 3),
multiplied by  $\Phi$(color singlet):
\begin{equation}
     \Phi_{JMTT_z, \:\xi}^{\rm (B)}  = \: \Phi \mbox{(color singlet)}
     \sum_{\gamma=1}^{3}
    \sum_{c \ne \gamma  }
  \Phi_{JMTT_z, \:\xi}^{(\gamma, c)}  \;
.
\label{waveb}
\end{equation}

The total wave function is given by a sum of these
 basis functions:
\begin{equation}
\Psi _{JMTT_z}=\sum_\xi  A_{\: \xi}\;  \Phi_{JMTT_z, \;\xi}^{\rm (A)}
+ \sum_{\xi^{\,'}}  B_{\, \xi^{\,'} }\;
\Phi_{JMTT_z, \;\xi^{\,'} }^{\rm (B)}
\; .
\label{total2}
\end{equation}
This is the most general form of the totally antisymmetric
three-quark wave functions of
$\Lambda, \Sigma, \Sigma^*,
\Xi$ and $\Xi^*$ (and their spatially excited states).
The A-type basis functions are useful for describing the correlations
between the quarks having the same isospin (strangeness)
whereas the B-type ones are effective for the
correlations between the quarks which have different
isospins (cf. Fig. 3).
The coefficients $A_{\:\xi}$ and $B_{\xi^{\,'}}$ are determined
by the variational principle.

It is to be noted that, 
when calculating the energy $E$ and the
coefficients $A_\xi$ and $B_{\xi'}$, use of 
the channel $\gamma=1$  alone in  Eq.~(\ref{waveb})
and (\ref{total2}) 
is sufficient since the strong 
interaction does not mix the configurations having
different $\gamma$.  But, use of the full wave function
with $\gamma=1-3$ (replacing the so-obtained
$A_\xi$ by $A_\xi/\sqrt{3}$ and $B_{\xi'}$ by
$B_{\xi'}/\sqrt{3}$) is necessary in the calculation of
the weak decay matrix elements.


\subsection{Spatial basis functions}

Our wave function
allows complicated admixture of spin-isospin states depending on
the quark-pair correlations.
For the spatial basis functions
$\phi _{n l m}({\bf r})$ and $\psi _{NLM}({\bf R})$,
we have to employ any functional form which satisfies the following
requirements:
i) The functions should be very suited
for describing both the short-range
correlations and the long-range
tail  behavior.
ii) Energy matrix elements should be calculated analytically and
easily
between the basis functions of the different
arrangement set of Jacobian coordinates.
iii) Non-linear parameters of the basis functions
can be searched  quickly.

To the authors' opinion, the most suitable  are
the Gaussian basis functions
with range parameters  chosen to lie in a
geometrical progression
\cite{Hiyama03,Kami88,Kame89,Hiyama95,Hiya96,Hiya00}: 
\begin{eqnarray}
      \phi_{n l m}({\bf r})
      &=&
      N_{n l} \; r^l \, e^{-(r/r_n)^2}
       Y_{l m}({\widehat {\bf r}})   ,
 \nonumber \\
      \psi_{NLM}({\bf R})
      &=&
       N_{NL}\: R^L \, e^{-(R/R_N)^2}
       Y_{LM}({\widehat {\bf R}})   ,
\label{gauss}
\end{eqnarray}
where $N_{n l}$ and $N_{NL}$ are normalization constants and
\begin{eqnarray}
      r_n
      &=&
      r_{\rm min} a^{n-1} \qquad \enspace
      (n=1 \sim n_{\rm max})  ,
\nonumber\\
      R_N
      &=&
      R_{\rm min} A^{N-1} \qquad
     (N \! =1 \sim N_{\rm max}).  
\label{range}
\end{eqnarray}
Successful applications of
the Gaussian expansion method
with the use of the above Gaussian basis functions
are seen in Refs. 9)-14)
for various three- and four-body systems.

\subsection{Matrix elements of the Weak Hamiltonian}

With the use of the above wave functions, the weak decay matrix elements,
$\left\langle {p\left| O_1^{NR}  \right|\Sigma ^+}
\right\rangle$,
$\left\langle {n\left| O_1^{NR}  \right|\Sigma ^0}
\right\rangle$,
$\left\langle {n\left| O_1^{NR} \right|\Lambda}
\right\rangle$,
are calculated in the following manner:
Firstly, we consider the operation of the operator Eq.~(\ref{opew})
on the hyperon wave function.
Since $s$ quark changes to $u$ quark and $d$ quark to $u$ quark
due to the weak interaction (Fig. 1),
the operation on the isospin part
of the $\Sigma^+$ wave function results in
\begin{eqnarray}
a_d^\dagger \,  a_u^\dagger \:(1 & - &
  \mbox{\boldmath $\sigma$}_u \cdot
  \mbox{\boldmath $\sigma$}_s )
\, \,
\delta({\bf x}_u - {\bf x}_s)
 \, \, a_u  a_s
\: | \:  \eta_{\frac{1}{2} \frac{1}{2}}(\alpha)
  \eta_{\frac{1}{2} \frac{1}{2}}(\beta)
  \eta_{00}(\gamma) \: \rangle \nonumber \\
  & = &
 (1 -
  \mbox{\boldmath $\sigma$}_\alpha \cdot
  \mbox{\boldmath $\sigma$}_\gamma
   ) \: \delta({\bf x}_\alpha - {\bf x}_\gamma)
\:|\:  \eta_{\frac{1}{2} -\frac{1}{2}}(\alpha)
  \eta_{\frac{1}{2}  \frac{1}{2}}(\beta)
  \eta_{\frac{1}{2}  \frac{1}{2}}(\gamma) \: \rangle \nonumber \\
 &+&
 (1 -
  \mbox{\boldmath $\sigma$}_\beta \cdot
  \mbox{\boldmath $\sigma$}_\gamma
   ) \: \delta({\bf x}_\beta - {\bf x}_\gamma)
\:|\: \eta_{\frac{1}{2} \frac{1}{2}}(\alpha)
  \eta_{\frac{1}{2}  -\frac{1}{2}}(\beta)
  \eta_{\frac{1}{2}  \frac{1}{2}}(\gamma) \: \rangle \;.
\end{eqnarray}
The same operation on the isospin part of the $\Lambda$ wave function
gives
\begin{eqnarray}
a_d^\dagger \,  a_u^\dagger \:(1 & - &
  \mbox{\boldmath $\sigma$}_u \cdot
  \mbox{\boldmath $\sigma$}_s )
\, \,
\delta({\bf x}_u - {\bf x}_s) \nonumber \\
&& \times  \, \, a_u  a_s
\: | \frac{1}{{\sqrt 2}}
\left[ \:  \eta_{\frac{1}{2}  \frac{1}{2}}(\alpha)
           \eta_{\frac{1}{2} -\frac{1}{2}}(\beta)
        - \eta_{\frac{1}{2}  -\frac{1}{2}}(\alpha)
          \eta_{\frac{1}{2}   \frac{1}{2}}(\beta) \right]
  \eta_{00}(\gamma) \: \rangle \nonumber \\
  & = &
 \frac{1}{{\sqrt 2}}(1 -
  \mbox{\boldmath $\sigma$}_\alpha \cdot
  \mbox{\boldmath $\sigma$}_\gamma
   ) \: \delta({\bf x}_\alpha - {\bf x}_\gamma)
\:|\:  \eta_{\frac{1}{2} -\frac{1}{2}}(\alpha)
       \eta_{\frac{1}{2} -\frac{1}{2}}(\beta)
       \eta_{\frac{1}{2}  \frac{1}{2}}(\gamma) \: \rangle \nonumber \\
 &+&
 \frac{1}{{\sqrt 2}}(1 -
  \mbox{\boldmath $\sigma$}_\beta \cdot
  \mbox{\boldmath $\sigma$}_\gamma
   ) \: \delta({\bf x}_\beta - {\bf x}_\gamma)
\:|\: \eta_{\frac{1}{2} -\frac{1}{2}}(\alpha)
      \eta_{\frac{1}{2}  -\frac{1}{2}}(\beta)
      \eta_{\frac{1}{2}  \frac{1}{2}}(\gamma) \: \rangle \;.
\end{eqnarray}
Overlap between the proton wave function
and the so-operated $\Sigma^+$ wave function,
multiplied by $G_F$sin$\theta$cos$\theta(c_1-c_2)/{\sqrt 2}$,
gives the amplitude
$\left\langle {p\left|{\cal H}_{\rm pc}  \right|\Sigma ^+}
\right\rangle$, and similarly for
$\left\langle {n\left|{\cal H}_{\rm pc}  \right|\Lambda}\right\rangle$.
From a simple relation between Clebsch-Gordan
coefficients, we have
$\left\langle {n\left|{\cal H}_{\rm pc}  \right|\Sigma ^0}
\right\rangle=$
$\left\langle {p\left|{\cal H}_{\rm pc}  \right|\Sigma ^+}
\right\rangle/{\sqrt 2}$.

\section{Mass spectrum and structure of the baryons}

We shall fix the model parameters $K, \beta$ and $C_{ss}/m_u^2$ so
as to reproduce the baryon  masses and the charge radius. The
experimentally measured proton charge radius includes
contributions from both valence quark core part and its meson
clouds. It is reasonable to subtract the vector meson dominance
contribution from the data of the proton electric charge radius
$(0.86 {\rm fm})^2$ to obtain the valence quark core radius
$\langle r^2_p \rangle _{\rm core}$. From this analysis, we
extract $\langle r^2_p \rangle  _{\rm core} \sim (0.6  {\rm
fm})^2$. By searching the parameters within a reasonable range, we
obtain the parameters $K=0.007  {\rm GeV}^3$, $\beta = 0.55$  fm
and $C_{SS} / m_u^2 \simeq 1.10  {\rm GeV}$ which give $m(\Delta)-
m(N) \simeq 293$  MeV and $\langle r^2_p \rangle  _{\rm core} =
(0.60  {\rm fm})^2$. After this determination, there are no more
adjustable parameters in our calculation. With these parameters,
we obtain the neutron charge radius square $ \langle r^2_n \rangle
= -0.05  \mbox{fm}^{2}$, which is also consistent with the
experimental value $ \langle r^2_n \rangle  = -0.12
\mbox{fm}^{2}$, after taking into account contributions from the
meson cloud.

As for the angular momentum space and the Gaussian range
parameters of the basis functions, we examined that the following
case is good enough. Contribution of the orbital angular momenta
other than $l=L=I=0$ is  found to be negligible as long as the
calculation is made of the ground states of baryons without
non-central forces as in the present study. As an example, spin
$s$ and $S$ and Gaussian range parameters employed for the
case of nucleon are listed in Table I only for $l=L=I=0$. The same
range parameters are used commonly for any set of spins and
isospins of other baryons and are suitable enough to obtain good
convergence of the eigenenergies.

\begin{table}
\caption{Three-body angular-momentum
space and Gaussian range parameters (in fm)
for the ground state of nucleon $(J=1/2^+, T=1/2)$.}
\label{table:1}
\begin{center}
\begin{tabular}{ccccccccccc} \hline \hline
$l$   &$L$  &$I$ &$s$  &$S$  &$n_{\rm max}$ &$r_{\rm min}$ &$r_{\rm max}$
&$N_{\rm max}$ &$R_{\rm min}$ &$R_{\rm max}$   \\
\hline
0 &0 &0 &0 &$1/2$  &8  &0.1  &3.0 &8 &0.1 &3.0 \\
0 &0 &0 &1 &$1/2$  &8  &0.1  &3.0 &8 &0.1 &3.0 \\
\hline
\end{tabular}
\end{center}
\end{table}

We show first in Fig. 4 SU(3) baryon mass spectrum. The agreement
with the data encourages us to proceed with our approach. It could
be possible to adjust the model parameters or to elaborate the
form of the potential to get much improved spectrum, but the
present results are enough for our purpose here.

\begin{figure}
\epsfxsize = 12cm   
\centerline{\epsfbox{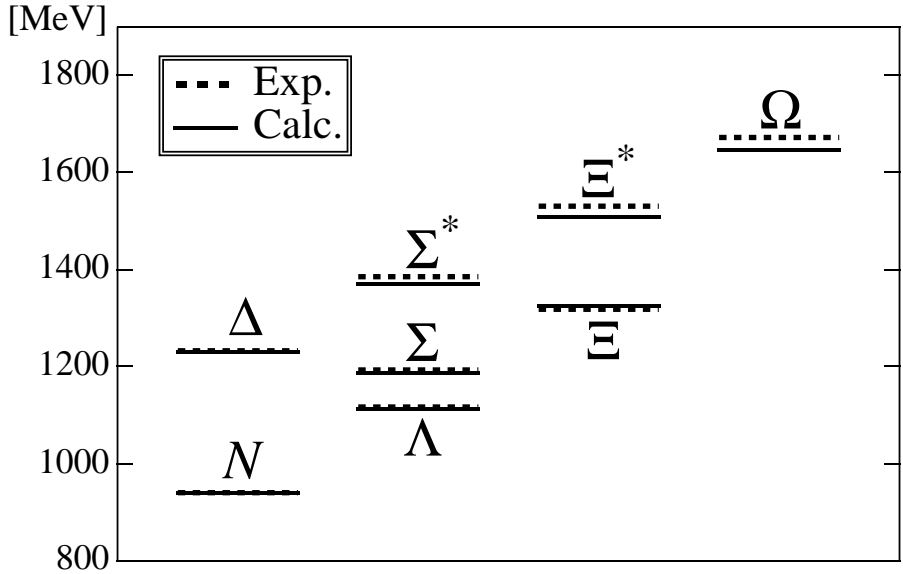}}
\vspace{0.1cm}
\caption{SU(3) baryon mass spectrum: Calculations are shown by the
solid lines, and experiments by the dashed ones. }
\label{mass}
\end{figure}

To clarify the effects of the attractive correlation
in the spin-0 quark pair on the nucleon structure,
we introduce the quark-pair correlation function.
\begin{eqnarray}
\rho^{(s)}( {\bf r}) = \langle \Psi_{\frac{1}{2} M \frac{1}{2}T_z} |
\delta({\bf x}_i  - {\bf x}_j - {\bf r})
P^{(s)}(ij)|\Psi_{\frac{1}{2} M \frac{1}{2}T_z} \rangle,
\end{eqnarray}
where {\bf r} stands for the distance between a quark pair and
the $P^{(s)}(ij)$ the projection operator of the quark-pair spin $s$;
in other words, the quantity
$\rho^{(s)}({\bf r})$ is the probability density to find
a spin-$s$ quark pair at the distance ${\bf r}$.
The density correlation function at origin $\rho^{(s)}(0)$ in the
$s=0$ case essentially fixes the value of the weak decay matrix element
in Eq.~(2.3).
$\rho^{(s)}({\bf r})$, which is independent of the angle
${\bf \hat{r}}$, is illustrated in Fig. 5.
In Fig. 5
we find
a large deviation between the $s=0$ and $s=1$ density distributions.
$\rho^{(s)}(0)$ in the $s=0$ case is about three times as
large as that of the $s=1$ case.
Similar tendency is seen in other baryons.
This provides a huge enhancement
for the $\Delta I=1/2$ weak decay amplitude.
\begin{figure}
\epsfxsize = 10cm   
\centerline{\epsfbox{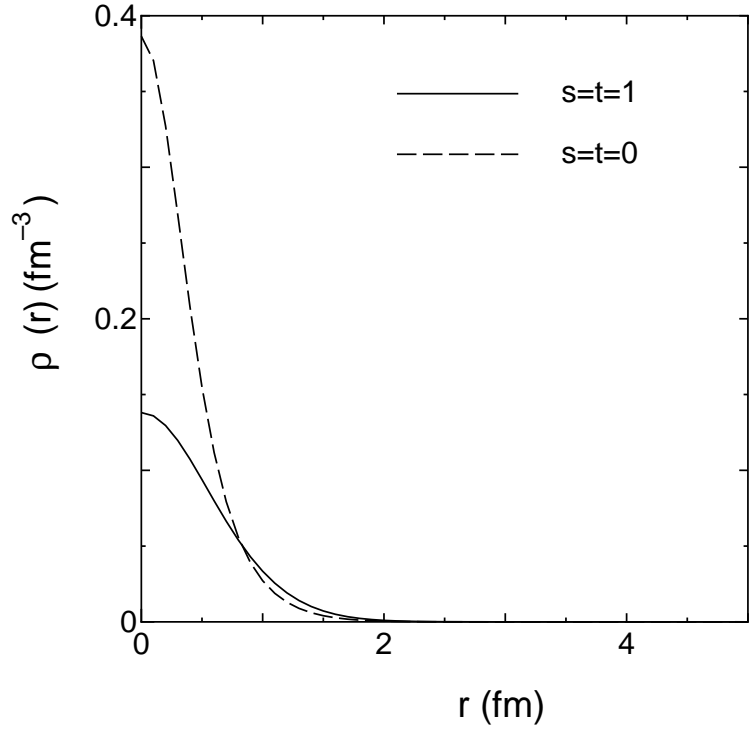}} \vspace{0.1cm} \caption{Quark-pair
density distribution $\rho^{(s)}({\bf r})$ with the spin-dependent
correlation.  Results for $s=t=1$ and $s=t=0$ pairs are depicted
by solid and dashed curves, respectively. } \label{fdensity1}
\end{figure}

For the nucleon, we calculate the r.m.s distance $\bar{r}^{(s)}$ and
$\bar{R}^{(s)}$ with respect to the Jacobian coordinates
${\bf r}$ and ${\bf R}$ by
$\bar{r}^{(s)}=[\int r^2 \rho^{(s)}({\bf r}) d{\bf r}]^{1/2}$
and similarly for $\bar{R}^{(s)}$. We get $\bar{r}^{(s=0)}=0.92$ fm
and $\bar{R}^{(s=0)}=0.97$ fm,  while
$\bar{r}^{(s=1)}=1.10$ fm and $\bar{R}^{(s=1)}=0.81$ fm.
Apparently, the $s=0$ correlation considerably
contacts $\bar{r}$ and extends $\bar{R}$ compared with
the $s=1$ case, but
the r.m.s matter radius
of the total system is almost
0.58 fm
both for the $s=0$ and $1$ cases.
It seems that the correlation is not so strong to form the
so called `diquark'-clustering
in the nucleon \cite{Diquark}.

\section{Numerical results for weak decay amplitudes}

\begin{wraptable}{r}{\halftext}
\caption{ Matrix elements of the operator, ${O}_1^{NR}$ (in
$10^{-2}{\rm GeV}^3$) with and without  the spin-dependent
attraction $V_s$.} \label{table:2}
\begin{center}
\begin{tabular}{ccc} \hline \hline
&with $V_s$   &without $V_s$    \\
\hline
$\langle n | O_1^{NR} | \Lambda \rangle$  &$-0.960$  &$-0.399$  \\
$\langle p | O_1^{NR}
| \Sigma^+ \rangle$  &$\enskip 2.760$  &$\enskip0.977$  \\
\hline
\end{tabular}
\end{center}
\end{wraptable}
With the parameters fixed in the previous section we calculate the
matrix elements of the weak Hamiltonian shown in Table II.
In the left column we show the matrix elements with the quark
correlation and those without the correlations in the right column.
In the absence of the correlation $V_{S} = 0$,
a ratio $\langle p | O_1^{NR} | \Sigma^+ \rangle /
\langle  n | O_1^{NR} | \Lambda \rangle = -2.45$ shows a perfect
agreement with the
SU(6) expectation $- \sqrt{6} \simeq - 2.4494\cdots$.
This agreement ensures the validity of our
numerical calculations (Not only the ratio but also the absolute value
have been examined).
In the realistic case with the spin-dependent force, one can observe the
substantial
enhancement of the matrix elements and the SU(6) breaking effects.

The non-leptonic weak transition parity-conserving
amplitudes are tabulated in Table III.
\begin{table}
\caption{$P$-wave non-leptonic weak transition amplitude (in $10^{-7}$
unit). In the second and forth columns, values in brackets show the
results without $V_s$.Empirical values are taken from ref. 1).}
\label{table:3}
\begin{center}
\begin{tabular}{ccccc} \hline \hline
Decay  &Pole  &others &Total  &Exp.   \\
\hline
$\Sigma^+ \rightarrow p \pi^0$  &$\enskip 23.23 (13.45)$  &$\enskip2.05$
&$\enskip 25.28 (15.50)$  &$\enskip26.6$ \\
$\Sigma^+ \rightarrow n \pi^+$  &$\enskip 40.9 (8.21)$  &$\enskip0.00$
&$\enskip 40.9 (8.21)$  &$\enskip42.2$ \\
$\Lambda \rightarrow n \pi^0$  &$-5.29 (-3.54)$  &$-5.02$
&$-10.31 (-8.56)$  &$-15.8$ \\
\hline
\end{tabular}
\end{center}
\end{table}
We show the pole contributions only in the second column and
additional factorization and penguin contributions, taken from
Ref. 4), are in the third column.  Then, we show the
total decay amplitudes in the forth column to be compared with the
experiments. It is worth noting that the $\Sigma ^+ \to n \pi^+$
parity conserving decay process is completely dominated by the
baryon pole diagrams without any factorization or penguin
contributions. This fact indicates that the $\Sigma ^+ \to n
\pi^+$ P-wave amplitude is the most appropriate observable to
probe the quark-quark correlation in the baryons. We find a good
agreement for $\Sigma \to N \pi$ decays, while the pole
contribution of the $\Lambda \to N \pi$ is not enough. In general,
our calculations reasonably reproduce the magnitudes of the pionic
hyperon decay amplitudes. However, there is a large cancellation
between the first and the second terms in Eq.~(\ref{lambdapole}),
which makes the resulting $\Lambda \to N \pi$ amplitude small.
Hence, the pole contribution to $\Lambda \to N \pi$ process
strongly depends on the values of $G_{\Lambda \Sigma }^{\pi 0},
G_{nn}^{\pi 0}$ as well as the weak matrix elements. Slight
variation of $G_{B \; B'}^{\pi ^a}$ $ \sim 10 \%$ yields about $40
\%$ modification of the $\Lambda \to N \pi$ pole contribution,
whereas the decay amplitudes for $\Sigma \to N + \pi^a$ are
essentially unchanged. The $\Lambda$ decay amplitudes are also
rather sensitive to the detailed shape of the potential.

We have obtained the agreement 
when we take the Gaussian size parameter $\beta
= 0.5 \sim 0.6$ fm for the spin-dependent potential.
If we choose smaller values of the size parameter $\beta$,
{\it e.g.}~0.2 fm,
the weak transition matrix element is enhanced and overestimates the
experimental data by order of magnitudes.
In this work, we have retained the non-relativistic kinematics so far.
Since the mass of the constituent quark is comparable with the kinetic
energy, we should
consider relativistic corrections to both the wave function and
the calculation of the weak decay matrix elements.
Inclusion of such relativistic corrections also changes the value of
size parameter $\beta$.
The study of such a dependence is  in progress.

In this paper, we concentrate on the calculation of the
parity-conserving P-wave amplitudes, since they are
particularly sensitive to the correlation strength.
On the other hand, for the parity violating S-wave amplitudes,
contributions from commutator, factorization, penguins and the
negative parity baryon pole terms are comparable.
This fact suggests that the S-wave amplitudes are not suitable to
quantify the quark-quark correlation.
Nevertheless, it is possible to calculate the S-wave
amplitude with our wave functions.
Results for the commutator contributions amplitudes are also consistent
with the data after including
the penguin, factorization and negative baryon pole
contributions. Such a study will be discussed in forthcoming paper.

\section{SU(6) symmetry breaking effects on the magnetic moment}

In the previous sections we have evaluated the pionic weak transition
amplitudes as well as the mass spectrum.  Due to the strong spin-dependent
attraction
our wave function enhances the weak decay amplitudes and
violates the naive SU(6) spin-flavor symmetry.
As we have shown in Table II,
the ratio of the matrix elements of the weak Hamiltonian
$\langle p | O_1^{NR} | \Sigma^+ \rangle /
\langle  n | O_1^{NR} | \Lambda \rangle$ shows a clear evidence for the SU(6)
breaking.
This ratio becomes $ -2.45$ without the correlation in a agreement with
the SU(6) result $-\sqrt{6}$, while
$-3.02$ with the quark correlation.
Therefore, we estimate the size of the SU(6) breaking effects
to be about $20 \%$, which is significant.

On the other hand,
it is historically known that the light baryon magnetic moments are well
reproduced
in the naive quark model by virtue of the SU(6) spin-flavor symmetry.
Hence, it is also interesting and important to estimate
effects of the SU(6) breaking by calculating the baryon magnetic
moments.

Neglecting contributions from the quark orbital angular momentum,
the operator for the baryon magnetic moment is simply given by
\begin{eqnarray}
\vec \mu_{mag}&=& \sum\limits_{i}\; \mu_i \; \sigma^z_i \; , \nonumber \\
&=& \sum\limits_{i}\; \frac{e}{2 m_i } Q_i \; \sigma^z_i \; ,
\label{magnetic}
\end{eqnarray}
where $Q_i$ is the electric charge operator of the $i$-flavor quark,
and $i$ runs over $i=1\sim 3$.
The magnetic moment of the baryon $B$ is obtained by directly calculating
the matrix element
\begin{eqnarray}
\mu_B =  \langle B | \vec \mu_{mag} | B \rangle
\end{eqnarray}
in terms of our wave functions.
This quantity certainly depends on the internal spin-flavor structure of the
 baryon through $m_i, Q_i$ and $\sigma^z_i$.
%
%

\begin{table}
\caption{Baryon magnetic moments}
\label{table:4}
\begin{center}
\begin{tabular}{cccc} \hline \hline
  &with $V_s$  &without $V_s$    &Exp.   \\
\hline
$\mu_p$  &$\enskip2.78$  &$\enskip2.84$  &$\enskip2.792847$ \\
$\mu_n$  &$-1.83$  &$-1.90$  &$-1.913042$ \\
$\mu_\Lambda$  &$-0.602$  &$-0.613$  &$-0.613 \pm 0.004$ \\
$\mu_{\Sigma^+}$  &$\enskip2.69$  &$\enskip2.73$
&$\enskip2.458 \pm 0.010$ \\
$\mu_{\Sigma^-}$  &$-1.05$  &$-1.06$  &$-1.160 \pm 0.025$ \\
$\mu_{\Sigma^0}$  &$\enskip0.817$  &$\enskip0.836$  &$-$ \\
$\mu_{\Xi^0}$  &$-1.410$  &$-1.449$  &$-1.250 \pm 0.014$ \\
$\mu_{\Xi^-}$  &$-0.507$  &$-0.502$  &$-0.6507 \pm 0.0025$ \\
$\mu_\omega$  &$-1.84$  &$-1.84$  &$-2.02 \pm 0.05$ \\
\hline
\end{tabular}
\end{center}
\end{table}

Our results are shown in Table IV.
In the first column, we show the results with the spin-dependent force,
and those without the correlations in the second column which
are the same as the naive SU(6) results.
It is manifest that the results are almost unchanged even after
introducing the spin-dependent correlations.  The differences are of
order of a few $\%$ in any cases.


These results tell us that
the global spin-flavor structure of quark wave functions (integrated over
volume) is quite insensitive to the existence of the quark-quark
correlations.
As a result, the success of the SU(6) symmetry for bulk baryon
properties seems to be maintained.
On the other hand, the introduction of the quark correlation
modifies the local structure of the
quark wave function substantially shown in Fig. 5, and thus enhances
the weak decay matrix elements.

The violation of the SU(6) spin-flavor symmetry in the local
quark distributions can be found in other
experiments, namely, the deep inelastic lepton scattering off the
nucleon.
Measured ratio of the momentum distribution functions
$d(x) / u(x)$ at $x \sim 1$ tends to 0 in contrast to the SU(6) value
$1/2$.
This is one of the examples that demonstrates the breakdown of the
{\em local} SU(6) spin-flavor symmetry.  We also note that
such a flavor dependence of the quark distribution functions is
also explained by considering the quark-quark correlation in the
$s=0$ channel\cite{Suzukisf}.


\section{Summary}

In conclusion, we have studied the role of the spin-dependent
quark-quark correlation in the baryon structure. We have pointed
out that the non-leptonic weak transition of the hyperon is an
unique quantity to investigate the quark correlation in the spin-0
channel. In particular, $\Sigma^+ \to n \pi^+$ P-wave decay is
free from the factorization or penguin contributions as shown in
Table III, and thus serves a severe  constraint on the
quantitative understanding of the spin-0 correlation. In order to
demonstrate its importance for the weak transition, we have
introduced the non-relativistic constituent quark model with the
spin-dependent attraction. We have solved three-body problem
rigorously using Gaussian expansion method
\cite{Hiyama03,Kami88,Kame89,Hiyama95,Hiya96,Hiya00}. Such a
spin-dependent interaction may originate from the non-perturbative
QCD dynamics. Results for static baryon properties reasonably
agree with the empirical values.

We have developed the procedure to calculate the matrix elements of the
weak Hamiltonian with a number of the rearrangement channels
as discussed in section 3.
Same technique can be applied to the strangeness
$-2$-system, $\Xi$ hyperons, and such a study is in progress.
Calculated transition amplitudes without the spin-dependent quark
correlation agree with the SU(6) predictions perfectly.  This result
guarantees the accuracy of our numerical calculation.
Resulting parity-conserving weak transition amplitudes are consistent
with the experiments, when the quark correlation is turned on.
At the present, quantitative description of the non-leptonic
weak hyperon decay is still difficult in any models of the
baryons\cite{model,QSRw,Skyrme}.
Naive chiral perturbation theory cannot explain the parity-conserving
and violating amplitudes simultaneously, and convergence of the chiral
expansion seems to be worse\cite{ChPT}.
Here, we present a possible improvement of this long standing problem
in the framework of the constituent quark model, but still have several
things to work out.  For example, the $\Lambda \to N \pi$ transition
amplitudes are underestimated due to the large cancellation in the
pole formula, although they are strongly dependent on the
choice of the parameters.

Introduction of the quark correlation, which is strong enough to explain
the non-leptonic weak decay, modifies the valence quark structure of the
baryons considerably.
The distance between quarks in the spin-0 pair becomes shorter by
$20 \%$.
Although the modification is not so large to induce the diquark-clustering
of the nucleon, such a tendency is consistent with the several
phenomenological studies\cite{diq-exp,Dosch}.

Although the spin-dependent correlation certainly violates the SU(6)
spin-flavor symmetry for the wave functions,
we have found that the
baryon magnetic moments are almost unchanged compared with the prediction
of the naive SU(6) model, as demonstrated in Table IV.
This is because the magnetic moments are quite insensitive to the
variation of the wave function at short distances $r, R\sim 0$.
There is a historical argument that the SU(6) symmetry assumption in the
quark model is indispensable to keep the impressive agreement of the
magnetic moment with the data.  However, it is now evident that,
even after including the SU(6) spin-flavor breaking effects on the
wave function, one can obtain quite reasonable values for the baryon magnetic
moments.

Recently, strong evidence for a new five-quark baryon 
state $\Theta^+(1540)$ (known as a pentaquark) 
has been reported by several groups 
\cite{LEPS,DIANA,CLAS}.
In Ref. 27), it is suggested
that the baryon is a bound state of four quarks and
an antiquark, containing two highly correlated $ud$ pair.
The quark-quark correlation potential
obtained in the present work 
to explain the weak decay of hyperons as well as
the baryon mass spectrum would be useful
in the study of the structure of the five-quark baryon.



\section*{Acknowledgements}

E. H. is supported by the Grant-in-Aids for the Scientific
Research from the Ministry of Education, Science, Sports and
Culture.

\appendix
\section{} 


We give some formulae to calculate the weak decay amplitudes.
First we summarize the expressions for the baryon pole contributions to
the parity conserving $\Lambda$ and $\Sigma$ weak decay.
\begin{eqnarray}
B (\Lambda \to n \pi^0) &=&
{{M_N+M_\Lambda } \over {f_\pi }}\left[ {G_{nn}^{\pi 0}
{1 \over {M_\Lambda -M_N}}\left\langle {n\left| {\cal H}_{PC} \right|\Lambda
 }
 \right\rangle}\right. \nonumber \\
 && \left.
 \hspace{3cm}{
  +\left\langle {n\left| {\cal H}_{PC}
 \right|\Sigma ^0} \right\rangle
  {1 \over {M_N-M_\Sigma }}G_{\Lambda \Sigma }^{\pi 0}} \right]
\end{eqnarray}
\begin{eqnarray}
B (\Sigma^+ \to p \pi^0)& =&
{{M_N+M_\Sigma } \over {f_\pi }}\left[ {G_{pp}^{\pi 0}{1 \over {M_\Sigma
-M_N}}\left\langle {p\left| {\cal H}_{PC} \right|\Sigma ^+} \right\rangle
}\right. \nonumber \\
&& \left.
\hspace{3cm}{
+\left\langle {p\left|  {\cal H}_{PC} \right|\Sigma ^+}
\right\rangle {1 \over
{M_N-M_\Sigma }}G_{\Sigma + \Sigma +}^{\pi 0}} \right]
\end{eqnarray}
\begin{eqnarray}
\hspace{-0.8cm} B (\Sigma^+ \to n \pi^+) &=&
{{M_N+M_\Sigma } \over {f_\pi }}\left[ {G_{pn}^{\pi +}
{1 \over {M_\Sigma -M_N}}\left\langle {p\left|{\cal H}_{PC}
 \right|\Sigma ^+}
\right\rangle } \right. \nonumber \\
&&\left.
\hspace{-1.5cm}{+\left\langle {n\left| {\cal H}_{PC} \right|\Sigma ^0}
\right\rangle
{1 \over {M_N-M_\Sigma }}G_{\Sigma +\Sigma 0}^{\pi +}+
\left\langle {n\left| {\cal H}_{PC} \right|\Lambda } \right\rangle
{1 \over {M_N-M_\Lambda }}G_{\Sigma +\Lambda 0}^{\pi +}} \right]
\end{eqnarray}
where ${\cal H}_{PC}$ is the parity conserving part of the weak Hamiltonian;
\begin{eqnarray}
{\cal H}_{PC} &=& \frac{G_F \sin \theta \cos \theta }
{\sqrt{2}}
\left( c_1  \left[
(\bar d \gamma_\mu u) (\bar u \gamma^\mu   s)
+ (\bar d \gamma_\mu \gamma_5 u) (\bar u \gamma^\mu \gamma_5 s) \right]
\right. \nonumber \\
& &  \left. +  c_2
\left[ (\bar u \gamma_\mu u) (\bar d \gamma^\mu   s)
+ (\bar u  \gamma_\mu \gamma_5 u) (\bar d \gamma^\mu \gamma_5 s)
\right]  \right) \;,
\end{eqnarray}
where we use only the current-current operators $O_1,O_2$ which give
dominant contributions here.
We omit the penguin contributions, since we are interested in the
absolute magnitudes of the weak transition amplitudes.
%
%
We use the following parameters of the weak Hamiltonian,
$G_F  = 1.16639 \times 10^{-5}
\mbox{GeV}^{-5}$,
$\sin \theta= 0.219$,
$\cos \theta = 0.975$, and $f_{\pi} = 92$MeV.

As for the axial vector coupling constant, we adopt the
SU(3) Goldberger-Treiman relation following to the standard approach.
\begin{eqnarray}
&& G_{nn}^{\pi 0} = G_{pp}^{\pi 0} = \frac{f_\pi}{2 M_N} (f+d) g\\
&& G_{\Lambda \Sigma 0 }^{\pi 0} = G_{\Sigma +\Lambda 0}^{\pi +} =
\frac{2f_\pi } {\sqrt{3}(M_N  + M_\Lambda)} d \,  g \\
&& G_{\Sigma + \Sigma + }^{\pi 0}= - G_{\Sigma +\Sigma 0}^{\pi +} =
\frac{2f_\pi}{2 M_\Sigma} f \, g \\
&&G_{p n }^{\pi + } = \sqrt{2} \frac{f_\pi}{2 M_N} (f+d) g
\end{eqnarray}
where $g$ is the strong $\pi NN$ coupling constant, and $f,d$ the SU(3)
coupling with a condition $f+d =1$.
We use $f=0.38$ and $d=0.62$ to obtain the numerical results presented in
this paper.  As pointed out in the text, the pole contribution
$\Lambda  \to N \pi$ is very sensitive to the choice of $f,d$.

The matrix elements of the weak Hamiltonian must be evaluated with
calculated quark wave functions.  The isospin symmetry for
$u$ and $d$ quarks implies a relation
\begin{eqnarray}
\left\langle {n\left| {\cal H}_{\rm PC} \right|\Sigma ^0} \right\rangle  =
\frac{1}{\sqrt{2}} \left\langle {p\left| {\cal H}_{\rm PC} \right|\Sigma ^+}
\right\rangle \; \,
\end{eqnarray}
which exactly holds with our wave functions.

\end{document}